\documentclass[runningheads]{llncs}
\usepackage[T1]{fontenc}
\usepackage{underscore}

\usepackage{graphicx}
\usepackage[utf8]{inputenc}
\usepackage{textcomp}
\usepackage{amssymb, pifont} % for \checkmark and \ding added by AD 05th Jul
\usepackage{todonotes} % for \todo command added by AD 05th Jul
\usepackage{url} 
\usepackage{caption}
\usepackage{subcaption}
\usepackage{booktabs}
\begin{document}

\title{Keywords are not always the key: A metadata field analysis for natural language search on open data portals}
\titlerunning{Keywords are not always the key}
% If the paper title is too long for the running head, you can set
% an abbreviated paper title here
%

% \author{Anonymous Author(s)}
% \authorrunning{Anonymous}
% \institute{}

\author{Lisa-Yao Gan\inst{1,3}\orcidID{0009-0004-3099-1738} \and
Arunav Das\inst{2}\orcidID{0009-0008-9989-1718} \and
Johanna Walker\inst{2}\orcidID{0000-0002-5498-8670} \and 
Elena Simperl \inst{2,3}\orcidID{0000-0003-1722-947X}
} 

\authorrunning{Gan et al.}
% First names are abbreviated in the running head.
% If there are more than two authors, 'et al.' is used.
%
\institute{Technical University Munich, Arcisstrasse 21, 80333 Munich \and
King's College London, Strand, WC2R 2LS London \and
Institute for Advanced Study, Technical University Munich, Lichtenbergstrasse 2a, D-85748 Garching, Germany\\
\email{lisa.gan@tum.de}}

\maketitle              % typeset the header of the contribution
\begin{abstract}
Open data portals are essential for providing public access to open datasets. However, their search interfaces typically rely on keyword-based mechanisms and a narrow set of metadata fields. This design makes it difficult for users to find datasets using natural language queries. The problem is worsened by metadata that is often incomplete or inconsistent, especially when users lack familiarity with domain-specific terminology. In this paper, we examine how individual metadata fields affect the success of conversational dataset retrieval and whether LLMs can help bridge the gap between natural queries and structured metadata. We conduct a controlled ablation study using simulated natural language queries over real-world datasets to evaluate retrieval performance under various metadata configurations. We also compare existing content of the metadata field 'description' with LLM-generated content, exploring how different prompting strategies influence quality and impact on search outcomes. Our findings suggest that dataset descriptions play a central role in aligning with user intent, and that LLM-generated descriptions can support effective retrieval. These results highlight both the limitations of current metadata practices and the potential of generative models to improve dataset discoverability in open data portals.

\end{abstract}
\section{Introduction}

\begin{figure}[!htb]
    \centering
    \includegraphics[width=0.85\textwidth]{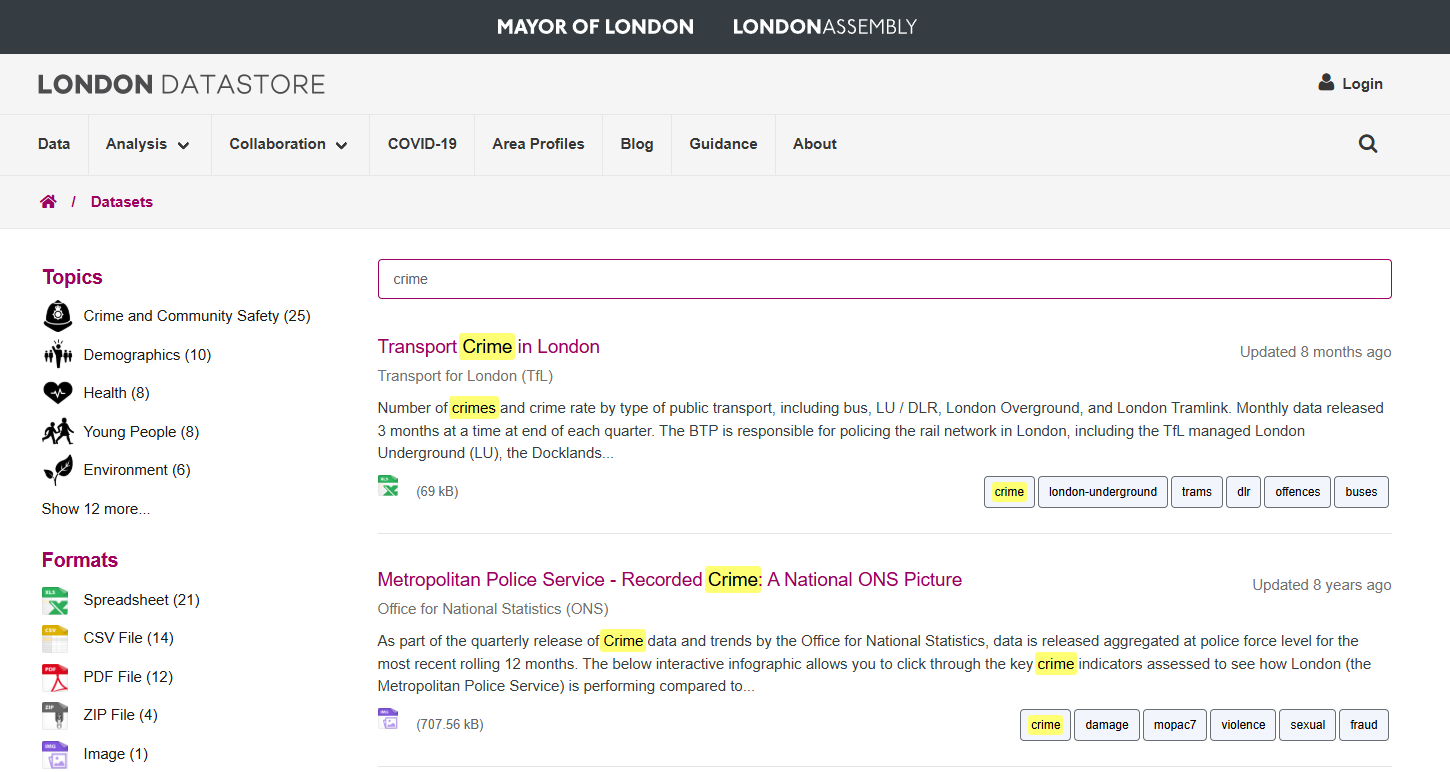}
    \caption{Example of keyword-based search interface in the London Datastore. The system uses exact term matching across limited metadata fields \cite{london_datastore}.}
    \label{fig:london_search}
\end{figure}

Open data portals play a crucial role in promoting transparency, civic engagement, and evidence-based policymaking by providing public access to government and institutional datasets \cite{10.1145/3657054.3657183,gurin}. Despite the increasing availability of such data, users often struggle to find datasets that match their information needs\cite{10.1145/3665939.3665959,10.1145/3025453.3025838}. Existing portals primarily support keyword-based search mechanisms that rely on exact term matching and predefined metadata fields such as title, keywords, and descriptions \cite{10.1145/3025453.3025838,10.1145/2964909,usercentered}. In addition, the quality of publisher-provided metadata is often low: it may be sparse, inconsistent, or missing for many datasets \cite{KOESTEN2021102562,Chapman_2019,dataliteracy}. These limitations hinder users in discovering datasets, as they may lack domain knowledge, use different terminology, or formulate queries in natural language \cite{searchlogs,Chapman_2019,article}. 
Figure~\ref{fig:london_search} illustrates this system in the London Datastore, where results are strictly based on lexical overlap. Filtering datasets is faciliatated via facets, displayed on the left hand side of the page. Users can filter the results based on topics, formats, geographical boundaries etc. No guidance or assistance is provided. Consequently, users iteratively refine and evaluate their search results, which makes traditional dataset search on open data portals a tedious and frustrating experience, especially for non-experts \cite{10.1145/3025453.3025838,zhao2024userexperiencedatasetsearch}.

At the same time, recent advancements in large language models (LLMs) have significantly changed the way users interact with digital systems: LLMs enable conversational interfaces that can interpret natural language questions, generate coherent answers, and understand user intent \cite{openai2024gpt4technicalreport,touvron2023llama2openfoundation}. This shift has increased attention on \textit{Conversational Information Retrieval} (CIR), where users search for data using natural language.  CIR systems offer promise in improving data discoverability by aligning search mechanisms with a more natural and fuller expression of need and intent \cite{gao2022neuralapproachesconversationalinformation,mo2024surveyconversationalsearch}.
While there is an active research direction that explores CIR for structured datasets and tables \cite{10.1145/3397271.3401205,odyssey,yu-etal-2019-cosql}, the transition to conversational interaction remains largely underexplored in practical, deployed open data portals. 

\subsection{Problem Statement}
Dataset retrieval in open data portals is still dominated by keyword-based search over static, human-authored metadata. As users increasingly rely on natural language to express their needs, this leads to a persistent mismatch between query intent and how data is indexed and retrieved \cite{KOESTEN2021102562,Chapman_2019,dataliteracy}. Metadata fields such as title, tags, and description are inconsistently populated and often misaligned with user vocabulary, making relevant datasets difficult to discover. Bridging this gap remains a central challenge in building more intuitive and effective dataset search systems.

\subsection{Research Questions}
This paper addresses the challenge of enabling effective conversational dataset discovery in structured open data portals. We focus on understanding the role of metadata fields and the effectiveness of LLM-generated content in supporting natural language queries. Our study is guided by the following research questions:

\begin{itemize}
    \item \textbf{RQ1:} Which metadata field (keyword, description, topic) is most important for effective dataset discovery using natural language queries?
    \item \textbf{RQ2:} What is the impact of LLM-generated descriptions on dataset retrieval performance compared to existing publisher-provided descriptions?
    \item \textbf{RQ3:} How do different prompting strategies influence the quality of dataset search outcomes?
\end{itemize}

\subsection{Contributions}
To address these questions, we conduct a controlled ablation study that isolates the impact of individual metadata fields on conversational dataset retrieval. We simulate real-world natural language queries and systematically evaluate retrieval performance under various metadata configurations.

Our central \textbf{hypothesis} is that \textit{the content of the metadata field 'description' is sufficient to support effective conversational dataset search}, and that \textit{LLM-generated description field content, derived directly from LLM interaction with the dataset, can outperform manually authored field content for the purpose of discovering relevant datasets}.

The main contributions of this paper are:

\begin{itemize}
    \item \textbf{Provision of the first metadata ablation study in conversational dataset search.} We systematically evaluate how individual metadata fields (title, description, topic, etc.) contribute to retrieval performance using natural language queries, addressing a gap in prior work which has focused more on search architectures or schema design than on empirical field-level analysis.

    \item \textbf{Evaluation of LLM-generated versus publisher-authored descriptions.} We compare the retrieval utility of LLM-generated metadata derived from raw dataset content with existing publisher-authored descriptions, providing insight into the value and limitations of generative metadata in real-world retrieval scenarios.

    \item \textbf{Prompting strategy analysis for query effectiveness.} Building on Walker et al.’s prompting styles \cite{walker2023promptingdatasetsdatadiscovery}, we assess how different query styles (requesting, describing, implying) interact with metadata configurations and affect retrieval outcomes, contributing to our understanding of user intent alignment in open data search systems.
\end{itemize}

This study provides empirical grounding for improving metadata practices, leveraging generative models, and designing user-centered retrieval systems in the context of open data.

\section{Related Work}
Effective dataset discovery remains a critical challenge across open data infrastructures. While the number of publicly accessible datasets continues to grow, search interfaces often rely on keyword-centric mechanisms that perform poorly when users express information needs in natural language~\cite{Chapman_2019,usability}. This section reviews prior work in two relevant areas: how users search for datasets in intent-driven settings and the role of metadata in dataset retrieval.

\subsection{How Users Search for Datasets in Natural Language}
Recent work in conversational dataset search has examined how users express information needs and how retrieval systems can better support intent-driven queries \cite{bonnet2025nls}.  Walker et al. \cite{walker2023promptingdatasetsdatadiscovery} studied user search interactions with open-domain chatbots. They categorize dataset search queries into three primary prompting strategies that reflect increasing levels of abstraction and ambiguity in user intent:

\begin{itemize}
    \item \textbf{Requesting:} This prompt type is \textit{goal-oriented and specific}, often including exact keywords or dataset names, formats, or even locations. Users employing this strategy tend to have a clear idea of what they’re looking for. These prompts are typically easy to interpret and retrieve against, making them most compatible with keyword-based retrieval systems. \\

    \item \textbf{Describing:} These prompts \textit{specify the features or structure} of the desired dataset, without naming a specific title or format. While less direct than requesting, they still express intent clearly, often listing attributes like granularity, timeframes, or coverage. \\

    \item \textbf{Implying:} The most \textit{open-ended and abstract} type, these prompts reveal a user’s broader goals or research interests rather than specifying a dataset directly. While in open-domain search, such prompts pose a greater challenge for retrieval models, particularly in inferring that a dataset would satisfy the user's intent, this problem does not arise in dataset-focused systems. 
\end{itemize}

These prompt types reflect different levels of user knowledge and task specificity. Walker et al. \cite{walker2023promptingdatasetsdatadiscovery} argue that effective conversational dataset search must accommodate all three forms, supporting clarification and dialogue where needed. Table~\ref{tab:prompting_styles} summarises the characteristics of these prompting styles, along with examples and their relative clarity of intent. Our study leverages this to generate diverse synthetic evaluation queries, ensuring that retrieval performance is tested across a realistic range of user expressions.

\begin{table}[!htb]
\caption{Prompting styles in dataset search (adapted from Walker et al.~\cite{walker2023promptingdatasetsdatadiscovery})}
\label{tab:prompting_styles}
\begin{tabular}{|p{2cm}|p{3.8cm}|p{4.5cm}|p{1.5cm}|}
\hline
\textbf{Prompt Type} & \textbf{Description} & \textbf{Example} & \textbf{Intent Clarity} \\
\hline
Requesting & Specific request for a dataset or format & \textit{“Find me a CSV of London air pollution in 2020.”} & High \\
Describing & Lists desired features without naming a dataset & \textit{“I need environmental data showing seasonal trends.”} & Medium \\
Implying & Expresses a goal or topic indirectly & \textit{“I'm exploring urban heat islands in Europe.”} & Low \\
\hline
\end{tabular}
\end{table}

\subsection{Metadata in Dataset Search}

Semantic search in on metadata-based dataset search is explored in recent academic work. Zhang and Balog\cite{10.1145/3441690} show that combining different metadata fields, such as column headers, captions, and surrounding context, into semantic representations can improve table search. Their work demonstrates that treating metadata as a multi-dimensional signal allows for more flexible and accurate matching. Other research focuses on cases where relevant information is spread across multiple tables. Chen et al.\cite{chen2025tableretrievalsolvedproblem} explore how to retrieve sets of tables that are not only relevant individually but also compatible with each other. They argue that understanding the structure of metadata—such as schema, column overlap, or key relationships—is essential for tasks like question answering over data lakes, where multiple sources often need to be combined.

Despite these advances, most open data portals still rely on relatively simple metadata-based search. CKAN-based platforms like Data.gov and the London Datastore, as well as Socrata-powered systems, use keyword matching on metadata fields \cite{ckan_search_features,socrata_catalog_faq,resourcesdatagov}. CKAN uses full-text engines like Apache Solr for filtering, relevance ranking, and partial string matching\cite{shahi2015apache}. Socrata improves match coverage through stemming\cite{socrata_catalog_faq}. Still, both systems depend on static, human-written metadata and exact term matching. The EU-wide portal \textit{data.europa.eu} aggregates national catalogues into a single CKAN-Solr index, but remains limited by the same metadata schema and keyword search approach\cite{dataeuropa_portal_architecture}. On a larger scale, Google Dataset Search indexes metadata embedded in schema.org annotations from web pages\cite{google_dataset_search,schema_org}. More recently, Google’s DataGemma project has tried to connect structured data with LLM-based assistants using sources like Data Commons\cite{radhakrishnan2024knowingaskbridging}. However, even these systems still rely on basic keyword logic rather than deeper semantic or structural understanding.

\section{Methodology}

Our study is grounded in the use of real-world metadata descriptions sourced from the London Datastore (LDS)\footnote{As of July 2025, the London Datastore was relaunched as the London Data Library (LDL), with some technical changes. This study was conducted prior to that transition, during the period it was known as the London Datastore.}. To investigate how different forms of metadata affect dataset discovery, we designed a methodology that combines dataset construction, metadata processing, modelling approaches, and evaluation metrics. 
Table~\ref{tab:rq_method_mapping} summarises how each research question (RQ) is addressed, linking each RQ to its corresponding methodological steps.

\begin{figure}[!htb]
    \centering
    \includegraphics[width=\linewidth]{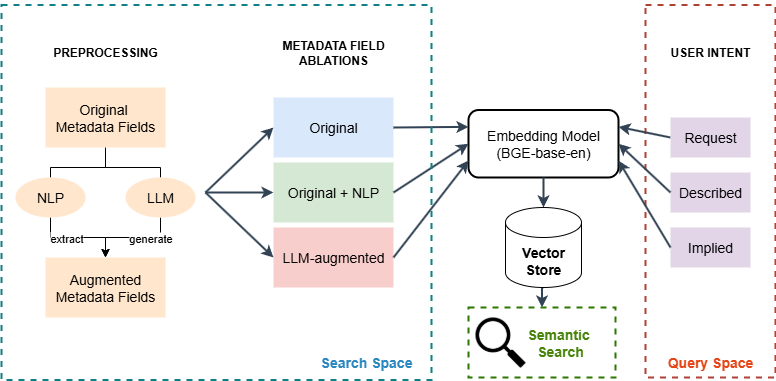}
    \caption{Overview of the data and augmentation pipeline used in our methodology}
    \label{fig:pipeline}
\end{figure}

\begin{table}[!htb]
\centering
\renewcommand{\arraystretch}{1.6}
\small
\caption{Overview of how each research question is addressed in the methodology.}
\label{tab:rq_method_mapping}
\begin{tabular}{|p{5cm}|p{7cm}|}
\hline
\textbf{Research Question} & \textbf{Approach} \\
\hline
\textbf{1. Which metadata field is most important for effective dataset discovery?} 
& Ablation over 13 metadata fields to measure individual contribution. \\
\textbf{2. What is the impact of LLM-generated descriptions on retrieval performance?} 
& Compare publisher vs. LLM-generated descriptions, individually and combined. \\
\textbf{3. How do prompting strategies influence dataset search outcomes?} 
& Evaluate search quality across request-, describe-, and imply-style queries. \\
\hline
\end{tabular}

\vspace{0.05cm}

% Second part: Tools and Evaluation (narrow first column)
\begin{tabular}{|p{2.5cm}|p{9.5cm}|}
\hline
\textbf{Tools} & Gemini 2.5 Flash, Gemma, BAAI/bge-base-en, FAISS, LSA, LDA, KeyBERT, spaCy NER \\
\textbf{Evaluation} & Top-1/3/5 Accuracy; Mean Reciprocal Rank (MRR) \\
\hline
\end{tabular}
\end{table}
Our retrieval methodology consists of two main components: the construction of a \textsc{Search Space} and the definition of a \textsc{Query Space}, which are illustrated in Figure~\ref{fig:pipeline}.

The \textsc{Search Space} is built from original metadata fields and two types of augmentation: (i) features extracted with traditional natural language processing (NLP) methods (e.g., keyword or topic extraction), and (ii) fields generated using LLMs. These are organized into three metadata configurations: (1) Original, (2) Original + NLP, and (3) LLM-augmented with several ablations defined within each configuration to test their relative contribution.

The \textsc{Query Space} captures user information needs through three query types: Request, Describe, and Imply. These represent increasing levels of abstraction and intent. Both metadata and queries are encoded using a shared embedding model and indexed in a vector database. Retrieval is then performed by comparing query embeddings to metadata embeddings, allowing us to evaluate how metadata richness and query style interact to influence search performance.

\subsection{Search Space Construction} \subsubsection{Characterising Metadata Completeness and Field Selection} 
\begin{figure}[!htb] \centering \begin{subfigure}[t]{0.49\textwidth} \centering \includegraphics[width=\textwidth]{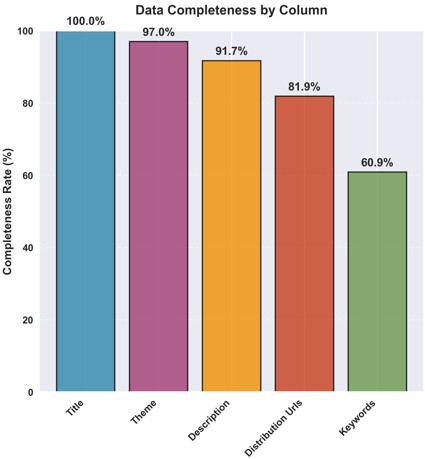} \caption{Field completeness across all 1,263 datasets.} \label{fig:metadata_completeness} \end{subfigure} \hfill \begin{subfigure}[t]{0.49\textwidth} \centering \raisebox{50pt}{ \includegraphics[width=\textwidth]{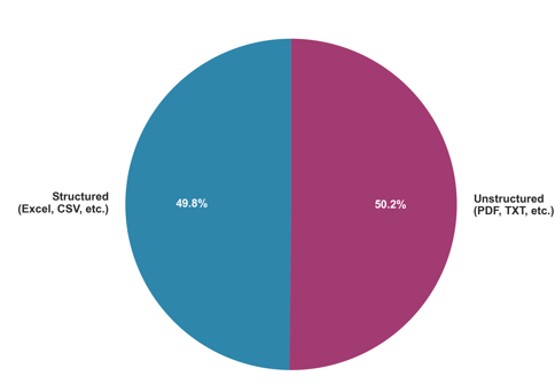} } \caption{Structured vs. unstructured file formats.} \label{fig:file_structure_pie} \end{subfigure} \caption{Metadata completeness and file format distribution across the London Datastore corpus.} \label{fig:metadata_coverage_combined} \end{figure}

The London Datastore corpus contains 1,263 datasets, but metadata quality is inconsistent: Our analysis shows that core fields such as \textit{title} (100\%), \textit{theme} (97.0\%), and \textit{description} (91.7\%) are consistently present, whereas only 81.9\% of datasets include a valid download link and just 60.9\% provide keywords. These gaps mirror findings from previous work on metadata quality \cite{Chapman_2019,dataliteracy,10.1145/2964909} and motivate our focus on enrichment strategies. Figure~\ref{fig:metadata_coverage_combined} illustrates both field completeness and file format distribution. As shown in Subfigure~\ref{fig:metadata_completeness}, descriptive fields are generally well covered, but keywords and distribution links remain inconsistent. Subfigure~\ref{fig:file_structure_pie} highlights that file formats are almost evenly divided between structured data (49.8\%, e.g., CSV and Excel) and unstructured formats (50.2\%, e.g., PDF and TXT). For downstream analysis, we restrict our study to the \textbf{255 datasets} that contain at least one structured file suitable for machine-readable processing. Within this filtered subset, our experiments focus on six metadata fields, summarised in Table~\ref{tab:metadata_fields}. Of these, \textit{dataset\_id} and \textit{title} are used only for dataset identification and human evaluation. The remaining fields \textit{description}, \textit{keywords} and \textit{topics} form the basis of our retrieval configurations. Throughout this paper, we use the term \textit{topic} to refer to the \textit{theme} field as defined in the London Datastore schema.  \renewcommand{\arraystretch}{1.3} % increase row spacing
\begin{table}[!htb]
\centering
\caption{Metadata fields used in the study and their role in retrieval}
\label{tab:metadata_fields}
\begin{tabular}{|p{2cm}| p{7cm}| p{2.5cm}|}
\hline
\textbf{Metadata Field} & \textbf{Description} & \textbf{Included in Ablation Study} \\
\hline
\textit{dataset\_id} & Unique identifier for the dataset, used for tracking and evaluation & -- \\
\textit{title} & Human-provided dataset title & -- \\
\textit{description} & Narrative description of the dataset content and purpose & \ding{51} \\
\textit{keywords} & Free-text keywords added by dataset publishers & \ding{51} \\
\textit{topics} & Predefined topical categories (e.g., "Transport", "Health") & \ding{51} \\
\hline
\end{tabular}
\end{table}
\renewcommand{\arraystretch}{1.0} % reset afterwards

\subsubsection{Metadata Enrichment}

Following the assessment of existing metadata completeness, we implemented two complementary approaches to enrich and complete missing fields: one based on traditional NLP techniques, and the other leveraging LLMs. We performed a manual review to spot-check whether the generated metadata reasonably matched the dataset content.

\paragraph{NLP-based Keyword and Topic Extraction}  

To enrich the often sparse publisher-provided metadata, we applied three complementary NLP techniques to dataset descriptions: Topic modelling was conducted using Latent Semantic Analysis (LSA) \cite{https://doi.org/10.1002/(SICI)1097-4571(199009)41:6<391::AID-ASI1>3.0.CO;2-9} and Latent Dirichlet Allocation (LDA) \cite{10.5555/944919.944937} to uncover hidden thematic structures and topic distributions in the text. For keyword extraction, we used KeyBERT \cite{grootendorst2020keybert}, which relies on BERT embeddings to identify words and phrases most semantically similar to the entire dataset description. This helps to produce keywords that better reflect dataset content and capture user phrasing even when exact wording differs. In addition, we employed Named Entity Recognition (NER) \cite{Honnibal_spaCy_Industrial-strength_Natural_2020} to identify and classify proper nouns and specific entities, such as organisations, locations, and technical terms, adding domain-specific vocabulary often missing from publisher-supplied keywords. Together, these methods produced a richer and more representative set of metadata, later integrated into retrieval experiments.

\paragraph{LLM-Generated Descriptions}  

In addition to NLP-based enrichment, we generated high-quality dataset descriptions using a large language model. This process involved two stages: structured prompt design and description generation.  

\textbf{Prompt design.} Metadata was first extracted from the London Datastore catalogue, including fields such as \textit{title}, \textit{description}, and download links. For datasets with valid machine-readable files (CSV, XLSX, XLS), we also incorporated structural context: column headers and a small sample of rows from the beginning and end of each dataset. To ensure clarity, headers and sample values were lightly sanitized (e.g., whitespace and newlines removed, excessively long strings truncated).  
These elements were then assembled into concise prompts that conveyed both the dataset’s structure and content. A simplified example is shown below:  

\begin{quote}
\small
\textit{Dataset \textbf{[title]} contains \textbf{[n]} records with column headers \textbf{[headers]}. Example records include: \textbf{[sample rows]}. Please generate a descriptive summary of the dataset (max. 350 words).}
\end{quote}

\textbf{Description generation.} Each prompt was submitted to the LLM with this instruction, and the resulting outputs were stored as a new \textit{llm\_description} field in the catalogue. To enable fair comparison with publisher-provided metadata, we subsequently applied the same NLP-based enrichment methods (topic modelling, keyword extraction, and NER) to these generated descriptions.  

\begin{figure}[!htb]
    \centering
    \includegraphics[width=\linewidth]{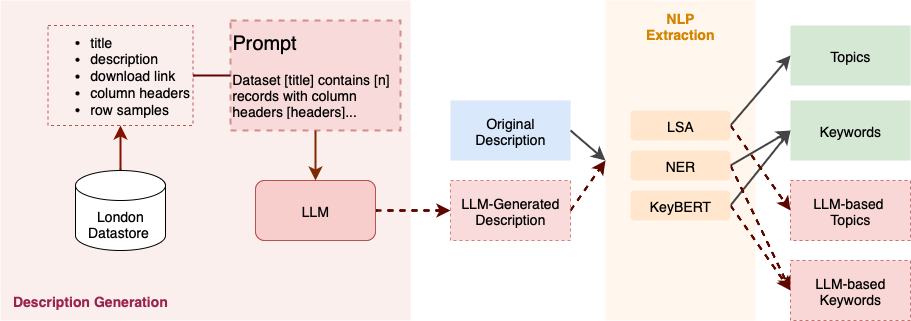}
    \caption{Overview of the metadata enrichment pipeline. Structured metadata and dataset samples are combined into prompts for an LLM, which generates synthetic descriptions. Both publisher-provided and LLM-generated descriptions are processed with NLP methods (LSA, NER, KeyBERT) to extract topics and keywords. Table~\ref{tab:generated_metadata_fields} lists the corresponding metadata fields produced at each stage.}

    \label{fig:metadata_enrichment_pipeline}
\end{figure}

To consolidate the fields used in our retrieval experiments, Table~\ref{tab:generated_metadata_fields} lists the original, NLP-derived, and LLM-derived metadata. Figure~\ref{fig:metadata_enrichment_pipeline} illustrates how these two enrichment strategies are combined into a unified pipeline.

\renewcommand{\arraystretch}{1.3} % adjust factor (1.2–1.3 usually looks good)
\begin{table}[!htb]
\centering
\caption{Metadata fields used in retrieval experiments, grouped by source.}
\label{tab:generated_metadata_fields}
\begin{tabular}{|p{3cm}|p{8.8cm}|}
\hline
\textbf{Source} & \textbf{Fields} \\
\hline
Original (LDS) & \texttt{lds\_title}, \texttt{lds\_description}, \texttt{lds\_keywords}, \texttt{lds\_topic} \\
NLP-derived & \texttt{lds\_desc\_keywords}, \texttt{lds\_desc\_topics} \\
LLM-derived & \texttt{llm\_prompt}, \texttt{llm\_description}, \texttt{llm\_desc\_keywords}, \texttt{llm\_desc\_topics} \\
\hline
\end{tabular}
\end{table}
\renewcommand{\arraystretch}{1.0} % reset back to default

\subsubsection{Ablation Configurations}
Having generated missing metadata fields using both NLP techniques and LLM-based methods, we now evaluate their contribution to dataset retrieval through a series of controlled ablations. In all ablation conditions, the fields \texttt{dataset\_id} and \texttt{title} are kept solely for identification and evaluation purposes. They are not included in the retrieval index or used as input for model inference. The different ablation configurations evaluated in our study are summarised in Table~\ref{tab:ablations}.

\renewcommand{\arraystretch}{1.3} % increase row height (1.1–1.3 looks good)
\begin{table}[!htb]
\centering
\caption{Ablation configurations used in retrieval experiments. Each configuration includes a subset of metadata fields, with variants derived from publisher-provided (Original), NLP-enriched, or LLM-enriched metadata.}
\label{tab:ablations}
\begin{tabular}{|p{3.4cm}|p{3.1cm}|p{2.7cm}|p{2.5cm}|}
\hline
\textbf{Configuration} & \textbf{Original} & \textbf{NLP} & \textbf{LLM} \\
\hline
Keywords + Topics & \texttt{key\_original} & \texttt{key\_nlp} & \texttt{key\_llm} \\
Description only & \texttt{desc\_original} & -- & \texttt{desc\_llm} \\
Full set (Description + Keywords + Topics) & \texttt{full\_original} & \texttt{full\_nlp} & \texttt{full\_llm} \\
Keywords only & \texttt{onlykey\_original} & \texttt{onlykey\_nlp} & \texttt{onlykey\_llm} \\
Topics only & \texttt{onlytopic\_original} & \texttt{onlytopic\_nlp} & \texttt{onlytopic\_llm} \\
\hline
\end{tabular}
\end{table}
\renewcommand{\arraystretch}{1.0} % reset to default

\subsection{Query Space Definition}
To evaluate the effectiveness of different metadata configurations in supporting natural language queries, we constructed a synthetic evaluation dataset. Inspired by the prompting styles identified by Walker et al.~\cite{walker2023promptingdatasetsdatadiscovery}, we generated three types of queries for each dataset in our corpus: \textit{requesting}, \textit{describing}, and \textit{implying}, which are summarized in table \ref{tab:prompting_styles}. To operationalize this, we used Gemini2.5 Flash to generate synthetic user queries, providing each model instance with the dataset title and ID as input context. This resulted in three diverse, natural-sounding queries per dataset.
Our final evaluation dataset therefore consists of 765 user queries in total.
Table~\ref{tab:user_query_example} presents an example of the three prompting styles applied to the dataset \textit{Police Force Strength}. 
\renewcommand{\arraystretch}{1.3} % increase row spacing
\begin{table}[!htb]
\centering
\caption{Example of generated user queries for a single dataset (Police Force Strength), based on Walker et al.'s prompting styles \cite{walker2023promptingdatasetsdatadiscovery}}
\label{tab:user_query_example}
\begin{tabular}{|p{2.5cm}|p{9.4cm}|} % widened prompt column for balance
\hline
\textbf{Prompt Type} & \textbf{User Query} \\
\hline
Requesting & Could you please help me locate the official dataset titled \textit{‘Police Force Strength’}? I'm interested in the latest figures available for police numbers across different areas. \\

Describing & I'm trying to find a dataset that details the current headcount and changes in personnel for various police forces. Do you have anything that provides comprehensive statistics on law enforcement strength? \\

Implying & What are the most recent statistics available regarding the total number of police officers currently serving in the country, and how has this changed over time? I'm curious about the staffing levels of our police force. \\
\hline
\end{tabular}
\end{table}
\renewcommand{\arraystretch}{1.0} % reset afterwards

\subsection{Retrieval and Evaluation Setup}

To evaluate the effectiveness of each metadata configuration, we implemented a dense retrieval pipeline. All metadata representations were encoded using the \texttt{BAAI/bge-base-en} embedding model \cite{bge_embedding}, a general-purpose English sentence transformer optimized for semantic similarity tasks. We used FAISS for efficient vector indexing and similarity search \cite{douze2025faisslibrary}. At query time, user queries were embedded using the same model, and cosine similarity was applied to retrieve the most relevant dataset representations. The top-$k$ results (with $k=5$) were then ranked by cosine distance between query and dataset embeddings.

Retrieval performance was assessed using standard information retrieval metrics \cite{voorhees-tice-2000-trec,bordes-etal-2014-question}:  
\begin{itemize}
    \item \textbf{Top-1 Accuracy:} Measures the proportion of queries for which the correct dataset appears as the top-ranked result.  
    \item \textbf{Top-3 and Top-5 Accuracy:} Evaluate whether the correct dataset appears within the top 3 or top 5 ranked results, respectively.  
    \item \textbf{Mean Reciprocal Rank (MRR):} A standard metric for ranking tasks that accounts for the position of the correct result, assigning higher weight to results retrieved earlier.  
\end{itemize}

These metrics provide a comprehensive view of retrieval quality across both exact and approximate matches, and are computed for each ablation condition and query style combination.

\section{Results}
We present the results of our ablation study evaluating how different metadata fields support natural language search. To maintain alignment with our research questions, we organize this section around the three RQs outlined earlier.

\subsection*{RQ1: Which metadata field is most important for effective dataset discovery using natural language queries?}

The results clearly show that \textit{descriptions} are the most effective metadata field for supporting natural language queries. The \textsc{desc\_llm} configuration, which uses only LLM-generated descriptions, achieves the highest performance across all metrics (MRR = 0.925), significantly outperforming the \textsc{desc\_original} configuration (MRR = 0.820). Even when evaluated in isolation, descriptions are more effective than any combination of keyword or topic metadata.

Keyword-based configurations (\textsc{key\_nlp}, \textsc{key\_llm}) perform moderately well but consistently lag behind descriptions. Topic-only fields perform the worst by a large margin, with MRR values below 0.1 in all variants. These findings highlight that rich descriptive metadata, particularly in narrative form, is essential for aligning with user intent in natural language queries.

It is worth noting that \textsc{desc\_original} already provides a strong human-authored baseline (MRR = 0.820, Hit@3 = 0.915), substantially outperforming all keyword- and topic-based configurations. This indicates that narrative descriptions, even when authored by publishers, capture user intent far better than structured metadata fields alone.

Table~\ref{tab:retrieval_results} summarises average retrieval performance across all queries for each configuration.

\begin{table}[!htb]
\centering
\caption{Retrieval performance across metadata ablation conditions. Scores reflect average retrieval accuracy across 765 natural language queries using a dense embedding-based retriever (BAAI/bge-base-en). Best results are highlighted in \textbf{bold}, while strong human-authored baselines (\textsc{desc\_original}) are also emphasized.}
\label{tab:retrieval_results}
\begin{tabular}{|l|c|c|c|c|}
\hline
\textbf{Ablation Condition} & \textbf{Hit@1} & \textbf{Hit@3} & \textbf{Hit@5} & \textbf{MRR} \\
\hline
\textsc{key\_original}       & 0.279 & 0.471 & 0.544 & 0.379 \\
\textsc{key\_nlp}            & 0.502 & 0.664 & 0.757 & 0.594 \\
\textsc{key\_llm}            & 0.495 & 0.680 & 0.753 & 0.594 \\
\textsc{desc\_original}      & \textbf{0.731} & \textbf{0.915} & \textbf{0.944} & \textbf{0.820} \\
\textsc{desc\_llm}           & \textbf{0.887} & \textbf{0.964} & \textbf{0.976} & \textbf{0.925} \\
\textsc{full\_original}      & 0.744 & 0.921 & 0.955 & 0.833 \\
\textsc{full\_nlp}           & 0.684 & 0.874 & 0.916 & 0.778 \\
\textsc{full\_llm}           & 0.835 & 0.940 & 0.956 & 0.887 \\
\textsc{onlykey\_original}   & 0.293 & 0.446 & 0.496 & 0.371 \\
\textsc{onlykey\_nlp}        & 0.492 & 0.676 & 0.756 & 0.592 \\
\textsc{onlykey\_llm}        & 0.499 & 0.669 & 0.759 & 0.597 \\
\textsc{onlytopic\_original} & 0.057 & 0.135 & 0.177 & 0.098 \\
\textsc{onlytopic\_nlp}      & 0.001 & 0.012 & 0.024 & 0.008 \\
\textsc{onlytopic\_llm}      & 0.009 & 0.027 & 0.038 & 0.019 \\
\hline
\end{tabular}
\end{table}

\subsection*{RQ2: What is the impact of LLM-generated descriptions on dataset retrieval performance compared to existing publisher-provided descriptions?}

We focus on configurations that isolate each source (\textsc{desc\_original} vs. \textsc{desc\_llm}) as well as those that combine descriptions with keyword/topic metadata (\textsc{full\_original} vs. \textsc{full\_llm}).

LLM-generated descriptions consistently outperform original ones, that were authored by publishers, across all setups. \textsc{desc\_llm} achieves an MRR of 0.925 compared to 0.820 for \textsc{desc\_original}. Similarly, \textsc{full\_llm} achieves 0.887 MRR, outperforming \textsc{full\_original} (0.833). 

Looking more closely, the gains are not only statistically consistent but also meaningful in relative terms: compared to \textsc{desc\_original}, \textsc{desc\_llm} improves Hit@1 by \textbf{+21.3\%} (0.731 → 0.887), Hit@3 by \textbf{+5.4\%} (0.915 → 0.964), Hit@5 by \textbf{+3.4\%} (0.944 → 0.976), and MRR by \textbf{+12.8\%} (0.820 → 0.925). This suggests that while publisher-authored descriptions are already a strong foundation, LLMs can provide additional semantic richness and better alignment with natural user phrasing, leading to consistently higher retrieval effectiveness.

\subsection*{RQ3: How do different user query types influence retrieval performance across metadata ablations?}

Performance varies considerably across query styles: As shown in Table~\ref{tab:retrieval_summary}, \textit{requesting} queries yield the highest retrieval scores across all configurations. In contrast, \textit{implying} queries result in the weakest performance, particularly when metadata is sparse.

For instance, the baseline \textsc{key\_original} configuration drops from 0.394 / 0.295 (MRR / Hit@1) for requesting queries to 0.351 / 0.246 for implying queries. In the worst case, \textsc{onlytopic\_llm} shows near-zero performance across all query types. Conversely, the best results come from \textsc{desc\_llm}, where scores remain consistently high even for vague queries (e.g., 0.906 / 0.858 for implying). Adding keyword and topic enrichment (\textsc{full\_llm}) boosts performance further for ambiguous inputs.

These results highlight how both metadata quality and query clarity shape retrieval outcomes and underline the need for richer, semantically informed metadata to support more natural, conversational search behavior.

\begin{table}[!htb]
\centering
\small
\caption{Retrieval performance (MRR / Hit@1) by query type for selected metadata configurations.}
\label{tab:retrieval_summary}
\begin{tabular}{|l|c|c|c|}
\hline
\textbf{Ablation} & \textbf{Requesting} & \textbf{Describing} & \textbf{Implying} \\
\hline
\textsc{key\_original}   & 0.394 / 0.295 & 0.392 / 0.295 & 0.351 / 0.246 \\
\textsc{desc\_llm}       & \textbf{0.962 / 0.943} & \textbf{0.907 / 0.861} & \textbf{0.906 / 0.858} \\
\textsc{full\_llm}       & 0.927 / 0.890 & 0.862 / 0.804 & 0.872 / 0.811 \\
\textsc{onlytopic\_llm}  & 0.024 / 0.014 & 0.017 / 0.007 & 0.017 / 0.007 \\
\hline
\end{tabular}
\end{table}

\subsection*{Summary}

Taken together, these results provide strong empirical support for our central hypothesis: that dataset descriptions, especially those generated by large language models, play a critical role in enabling effective natural language querying for dataset search. Across all metadata configurations and query types, LLM-generated descriptions consistently outperform original metadata. These findings point to a promising direction for data portals: augmenting or replacing sparse publisher metadata with semantically rich LLM-generated content.

\section{Discussion}

In this section, we reflect on the broader implications of our findings for the design, evaluation, and governance of conversational dataset search systems. 

\subsection*{LLMs as Metadata Enrichment Tools}

Our results confirm a growing trend observed in other domains: LLMs can generate metadata that is not only semantically rich but also more retrieval-effective than existing human-authored content (which is often inconsisistent and sparse). Across all configurations and query types, LLM-generated descriptions substantially outperformed original descriptions, keywords, and topics. This echoes findings in related work, where well-structured generation based on underlying data improves search relevance and interpretability \cite{10.1145/3441690}.

Unlike prior work that uses LLMs primarily for user-facing generation, our study demonstrates their backend utility in metadata generation. Given the high performance of \textsc{desc\_llm} and \textsc{full\_llm}, we argue that LLMs should be considered core components of metadata pipelines for open data portals, where manual curation is infeasible or in situations where metadata is gathered from a number of different sources, which inevitably introduces inconsistency.

\subsection*{Prioritising Description Metadata in Open Data Portals}

Traditionally, dataset portals have prioritised metadata fields such as keywords, formats, or predefined taxonomies (e.g., CKAN's topic structure). Our study challenges this emphasis. Despite widespread use, topic fields proved ineffective in conversational retrieval (MRR < 0.1 in most settings). These results support Koesten et al.'s critique of "disconnected" metadata, which fails to reflect user mental models and goals \cite{KOESTEN2021102562}.

Conversely, narrative descriptions, particularly when LLM generated, proved to be the most robust across varying query styles. This reinforces Chapman et al.'s call for semantically expressive metadata and aligns with Walker et al.'s classification of prompting strategies \cite{Chapman_2019,walker2023promptingdatasetsdatadiscovery}. In short, successful dataset discovery depends not on formal metadata fields alone, but on the richness and alignment of content with user language.

\subsection*{Implications for Interface and Infrastructure Design}

From a systems perspective, our findings offer a practical blueprint for building next-generation dataset portals. First, metadata enrichment pipelines using LLMs should become a standard feature of catalogue infrastructure. Second, retrieval engines should prioritise semantically dense fields (descriptions, column-level summaries) over sparse, taxonomic ones.. Third, search interfaces should be reimagined to support conversational interaction from the ground up. This involves not only adopting dense retrieval backends, as we used here, but also aligning front-end interfaces with real user workflows. Retrieval should not assume final-form queries; rather, systems should support progressive discovery based on incomplete or ambiguous inputs.

While not the focus of this paper, preliminary experiments suggest that retrieval performance is also influenced by the structure and length of user queries. Specifically, we observed that longer, more detailed queries tend to benefit more from rich narrative metadata such as LLM-generated descriptions, whereas short or keyword-like queries may align better with concise fields like titles or keywords. This suggests a need to explore how description-based metadata can support a wider search space, specifically, how LLM-generated descriptions can be optimized to improve retrieval for both short, keyword-style queries and longer, natural language ones.

\subsection*{Toward Standardized Machine-Generable Metadata}

Given the success of structured prompting in generating useful metadata, we propose that the data portal  community consider formalizing LLM-compatible metadata standards. Such standards might define prompt schemas (e.g., using column names and keywords), output formats (e.g., descriptive paragraphs or JSON metadata blocks), and evaluation criteria (e.g., alignment with schema.org or retrieval effectiveness). These would enable reproducibility across portals, improve metadata interoperability, and allow future systems to audit or improve generated content.

We envision a future in which data providers submit minimally structured metadata, and machine-generation pipelines produce standardized, rich metadata to populate portal interfaces, APIs, and downstream retrieval models.

\subsection*{The Missing Link Is the User}

A persistent challenge in dataset search research is the lack of access to real user interaction data. While we follow Walker et al. \cite{walker2023promptingdatasetsdatadiscovery} in using structured prompting as a principled proxy, the field as a whole still lacks large-scale, authentic user queries, session logs, and task outcomes needed to evaluate systems under realistic conditions.

This absence is not unique to our work. Koesten et al. \cite{KOESTEN2021102562} emphasize how poorly understood user needs are in open data portals, noting that users often do not know what to search for and engage in iterative query reformulation. As long as retrieval systems are evaluated without authentic user context, it will remain unclear how well they serve actual discovery tasks.

We therefore join prior work \cite{Chapman_2019,KACPRZAK201937} in calling for the creation of benchmark datasets containing real user queries, tasks, and satisfaction signals. These resources are crucial for defining what “successful” dataset discovery actually means in different use contexts.

\subsection*{Cross-Domain Relevance}

While our study focused on the London Datastore, the issues we address (metadata sparsity, intent misalignment, rigid search interfaces) are widespread across public and private data platforms. Scientific repositories, clinical data registries, and even internal enterprise data catalogues face similar challenges. Our approach in combining metadata ablation, structured prompting, and dense retrieval evaluation can readily be extended to other domains.

Ultimately, our findings suggest that metadata should not be seen as static infrastructure but as a dynamic, learnable surface for interaction. By treating metadata generation as an LLM-supported design problem, we can move toward truly user-centered dataset discovery across contexts.

\section{Future Work and Limitations}

While our findings offer compelling evidence for the effectiveness of metadata generated by LLMs in conversational dataset search, several limitations of scope, methodology, and generalizability must be acknowledged.

We emphasize that our comparisons were not made against carefully curated, high-quality human-authored metadata, but rather against the metadata currently available 'in the wild'. While this reflects the reality of open data portals, we acknowledge that our human-authored dataset will have necessarily included a number of poor descriptions. Our future work will explore evaluations against a subset of human-annotated “ideal” descriptions to more rigorously benchmark the strengths and limitations of LLM-based metadata.

\subsubsection*{Corpus Scope and Generalizability}

Our evaluation was conducted on a controlled set of 255 tabular datasets from the London Datastore. While this provided a manageable testbed for isolating metadata effects, it does not capture the scale, diversity, or complexity of larger open data ecosystems. Many real-world portals contain thousands of datasets across heterogeneous formats (e.g., PDFs, APIs, spatial data) and domains. A limited corpus may also inflate retrieval scores due to lower semantic overlap between documents. Future work should extend these evaluations to more diverse collections.

\subsubsection*{Metadata Schema Constraints}

We operated within the standard metadata fields available in common data portals. While this improves comparability and realism, it constrains the potential of more expressive or LLM-oriented metadata structures. As discussed, future work should explore redesigned metadata schemas that explicitly support LLM compatibility, conversational querying, and semantic richness.

\subsubsection*{Synthetic Queries and Missing User Context}

Although our evaluation used a principled classification of query styles from Walker et al.~\cite{walker2023promptingdatasetsdatadiscovery}, all user queries were synthetically generated. These prompts are structured, consistent, and lack the noise, ambiguity, and iterative reformulation common in real user behavior. As Koesten et al.~\cite{KOESTEN2021102562} and others have noted, user needs in dataset search are often underspecified and evolve over time. Without interaction logs or task-based evaluation provided by the London Datastore itself, our results cannot fully capture how well LLM-generated metadata supports authentic discovery scenarios. Future research must incorporate real conversational query data, user feedback, and success metrics to close this gap.

\subsubsection*{Retrieval Architecture Assumptions}

We employed a general-purpose dense retriever (BAAI/bge-base-en) to evaluate semantic alignment between queries and metadata fields. While this reflects current practices in neural retrieval, it does not explore the full design space of conversational IR architectures, such as query rewriting, reranking, or interactive clarification loops. Additionally, no retriever is neutral: our results may be influenced by how well the model embeds certain fields (e.g., long-form descriptions vs. sparse keywords). Tuning or training retrievers specifically for metadata structure or conversational prompts remains an open area.

\section{Conclusion}

This study examined how different metadata fields, and their enrichment via LLMs, affect conversational dataset retrieval. Through an ablation study on 255 datasets from the London Datastore and 765 natural language queries, we found that the \textit{description} field is the most critical for retrieval effectiveness, particularly when generated by LLMs. 

LLM-generated descriptions consistently outperformed publisher-authored ones, especially for vague or exploratory queries, highlighting their potential to improve metadata quality and align more closely with user intent. In contrast, traditional fields like keywords and topics showed limited utility in conversational search scenarios.

While limited in scope to a mid-sized tabular corpus, our findings point to a broader opportunity: integrating LLM-based metadata generation into open data infrastructure to support more natural, effective search. This work contributes practical guidance and empirical evidence for designing metadata pipelines and retrieval systems that reflect how users actually search in the age of conversational AI.

\begin{credits}
\subsubsection{\ackname}
Funded by the SIEMENS AG and the Technical University of Munich – Institute for Advanced Study, Germany and the King's College London NMES Partnerships Fund. We would like to acknowledge the support of the Data for London team.

\subsubsection{\discintname}
The authors have no competing interests to declare that are relevant to the content of this article.
\end{credits}

%
% ---- Bibliography ----
%
% BibTeX users should specify bibliography style 'splncs04'.
% References will then be sorted and formatted in the correct style.
%
\bibliographystyle{splncs04}
\bibliography{reference}
% %
% \begin{thebibliography}{8}
% \bibitem{ref_article1}
% Author, F.: Article title. Journal \textbf{2}(5), 99--110 (2016)

% \bibitem{ref_lncs1}
% Author, F., Author, S.: Title of a proceedings paper. In: Editor,
% F., Editor, S. (eds.) CONFERENCE 2016, LNCS, vol. 9999, pp. 1--13.
% Springer, Heidelberg (2016). \doi{10.10007/1234567890}

% \bibitem{ref_book1}
% Author, F., Author, S., Author, T.: Book title. 2nd edn. Publisher,
% Location (1999)

% \bibitem{ref_proc1}
% Author, A.-B.: Contribution title. In: 9th International Proceedings
% on Proceedings, pp. 1--2. Publisher, Location (2010)

% \bibitem{ref_url1}
% LNCS Homepage, \url{http://www.springer.com/lncs}, last accessed 2023/10/25
% \end{thebibliography}
\end{document}